\documentstyle[preprint,prb,aps]{revtex}

\begin{document}

\preprint{UNIFR/IPT-96}
\draft
\title{Friedel transition in a modified XY model}
\author{M. Dzierzawa, M. Zamora, D. Baeriswyl and X. Bagnoud}
\address{
Institut de Physique th\'{e}orique, Universit\'{e} de Fribourg, 
P\'{e}rolles\\
CH-1700 Fribourg, Switzerland}
\date{\today} 
\maketitle

\begin{abstract}

Weakly coupled superconducting layers are described by the three-dimensional
XY model with strong coupling in two directions and weak coupling in the
third direction. For the usual Josephson-type interplane coupling the
coherence between the layers is lost at the same temperature as that
within the layers. Thus a low-temperature layer decoupling due to
a proliferation of fluxons between planes, as proposed by Friedel,
does not occur in this case. However, for a modified interplane coupling
there are two phase transitions, one of a Kosterlitz-Thouless type
from a disordered high-temperature phase to an intermediate phase
with phase coherence only parallel to the layers, the second from this
effectively two-dimensional phase to a three-dimensional phase with
coherence in all directions and a finite "n-state" order parameter.
Thus we do find a "Friedel transition" for this special class of models.
\end{abstract}


\pagebreak

The Berezinskii-Kosterlitz-Thouless (BKT) transition \cite{Ber71,Kos73},
mediated by the unbinding of vortex pairs, has been clearly observed
in superfluid films \cite{Bis78}. More recently, nonlinear transport
experiments in layered high-temperature superconductors \cite{Art89} have
also shown typical signatures of vortex unbinding slightly below
the critical temperature $T_c$. This is surprising, since the
Josephson coupling between the layers renders the system three-dimensional,
in particular close to $T_c$. Specific heat experiments 
for YBCO single crystals indeed give
evidence for three-dimensional critical behavior \cite{Jun96}.

Some time ago Friedel \cite{Fri88} has argued that the interlayer coupling 
could be effectively suppressed  by a proliferation of Josephson vortex 
loops or fluxons between the layers. A simple
estimate for the fluxon energy suggests that the layers are decoupled at
a temperature $T^\ast < T_c$, so that there would be a two-dimensional
regime for $T^\ast < T < T_c$, with a BKT transition at $T_c$. 
Unfortunately, a closer look at the problem \cite{Fri89} indicates that this
``Friedel transition" does not occur below $T_c\;$  \cite{Kor90}.
Therefore the question arises why nevertheless in nonlinear 
transport experiments BKT signatures are 
clearly observed. A way out of this dilemma
has been offered by Jensen and Minnhagen \cite{Jen91}, 
who realized that the Lorentz force acting on the vortices can
overcome the interlayer confinement of vortex pairs. 

In this Letter we study again the possible existence of a two-dimensional
regime below the critical temperature, starting from the classical
XY model with strong intralayer coupling $J_\parallel$ and weak interlayer
coupling $J_\perp$. A simple criterion for the decoupling transition
together with Monte Carlo simulations shows that a layer decoupling below
$T_c$ does not occur, in agreement with previous studies \cite{Kor90}.
We attribute this negative result to a strong increase of the fluxon energy
as a function of temperature. We turn then to the question whether a decoupling
can be excluded on general grounds, by considering two modified XY models.
For the first model, representing a superlattice of high and low $T_c$
layers \cite{Tri94}, a decoupling seems to occur around the low $T_c$,
but a closer look shows that this apparent transition is in reality
a crossover from strong to weak interplane coherence. In the second
model a modified interlayer coupling is considered, allowing for $n$
equivalent phase differences for the superconducting order parameters
of adjacent layers \cite{Atthi}. In this case a decoupling at
$T^\ast < T_c$ is clearly observed for $n > 2$.

Incidentally, the problem is of more general relevance, as the question of
long-range coherence arises also in other
quantum systems. One can for instance ask whether a
disordered layered system of electrons, with a 
large difference between the masses for
the motion parallel and perpendicular to the layers, can be metallic in one and
insulating in another direction. Anderson has argued that this can happen, if 
electron-electron interactions are taken into account \cite{And91}.
A measure for quantum coherence in the case of electronic transport
is the Drude weight or charge stiffness, while in the context of
superfluidity the relevant quantity is the superfluid density $\rho_s$
(or the helicity modulus, in the language of the XY model).

We consider the classical XY model on a cubic lattice
\begin{equation}
H = - \sum_{i,\mu} J_\mu \cos (\varphi_i - \varphi_{i+\mu})\, ,
\label{1}
\end{equation}
where $\mu = x,y,z$, $J_x = J_y = J_\parallel$, $J_z = J_\perp$, and the phases
are restricted to $0 \le \varphi_i < 2\pi$. This model can be derived from the
anisotropic Ginzburg-Landau or the Lawrence-Doniach model by neglecting
both the fluctuations of the electromagnetic field and the amplitude
fluctuations of the order parameter. There are good arguments for 
doing this ,
although these degrees of freedom may become relevant within the critical
region \cite{Zum93,Bor94}. 
The helicity modulus $\Upsilon_\mu$ is defined 
as the second derivative
of the free energy with respect to a constant phase gradient in the direction
$\mu$, and can be written as
\begin{equation} 
\Upsilon_\mu   = 
\frac{J_\mu}{N} 
 \left<\sum_i  \cos (\varphi_i - \varphi_{i+\mu})\right> 
 - \frac{\beta J^2_\mu}{N} \left<\left( \sum_i \sin (\varphi_i -
\varphi_{i+\mu})\right)^2\right>  \,\, .
\label{2}
\end{equation}
According to Friedel's original suggestion \cite{Fri88}, for very weak
interlayer coupling  $\Upsilon_\perp$ would
vanish at a lower temperature than $\Upsilon_\parallel$,
leaving an intermediate temperature region of essentially
$2d$ character.

We present now a simple argument against a layer decoupling below
the critical temperature, by expanding $\Upsilon_\perp$ in powers
of $J_\perp$. The leading order coefficient
($\sim J_\perp^2$) turns out to be
the difference of two equal contributions, each of them given by
$S = \int d^2r \, c^2(r)$. The $2d$ correlation function
$c(r) = \langle\cos(\varphi_i - \varphi_{i+r}) \rangle$ decays exponentially
as a function of distance $r$ for $T > T_{KT}$.
Therefore the leading term in the expansion of $\Upsilon_\perp$
vanishes identically above the BKT transition (and the same can
be shown for all higher order terms).
For $T < T_{KT}$ $c(r) \sim r^{-\eta(T)}$, where 
$\eta(T) \stackrel{<}{\sim} 1/4\;$
\cite{Kos74,All88}, which implies that the quantity $S$ is infinite
below the BKT transition. We tentatively associate the temperature where 
$S$ diverges with
the transition from an effectively $2d$ phase of decoupled layers to a $3d$ 
phase with finite $\Upsilon_\perp$. According to this criterion the 
decoupling transition temperature $T^\ast$ coincides with the
critical temperature $T_c = T_{KT}$ in the limit $J_\perp \rightarrow 0$.

In order to gain a qualitative understanding of the temperature dependent
helicity modulus, we first use the renormalized harmonic
approximation (RHA), where the Hamiltonian  (\ref{1}) is replaced by an
effective harmonic term
\begin{equation}
\tilde{H} = \frac{1}{2} \sum_{i,\mu} \tilde{J}_\mu (\varphi_i -
\varphi_{i+\mu})^2\, 
\label{3}
\end{equation}
with variational parameters $\tilde{J}_\mu$. These parameters turn out to be
identical to the helicity moduli, $\tilde{J}_\mu = \Upsilon_\mu$. 
The anisotropy 
$\Upsilon_\perp/\Upsilon_\parallel$ diminishes 
with increasing temperature, but remains finite up to
the critical temperature, where both 
$\Upsilon_\perp$ and $\Upsilon_\parallel$ drop to zero
discontinuously (Fig.1).
We have also performed Monte Carlo simula\-tions using the
standard Metropolis algorithm. The results presented in Fig.1 
show that the jumps obtained in the RHA are artifacts
and that the helicity moduli  tend to zero continuously. 
We note the excellent agreement between the two 
methods at low temperatures, as expected. Our numerical results for
$J_\perp = 0.1 J_\parallel$ are consistent with a simultaneous
loss of coherence parallel and perpendicular to the layers at
$T_c \approx 1.33 J_\parallel$. Therefore the temperatures $T^\ast$
and $T_c$ coincide, in agreement with the simple criterion discussed above.

We focus now our attention on the role of fluxons
by considering an approximate
version  of the XY model, where the nonlinearity is retained only 
for the interlayer coupling,
\begin{equation}
H = \frac{J_\parallel}{2} \sum_{i,\mu=x,y} (\varphi_i - \varphi_{i+\mu})^2
- J_\perp \sum_{i} \cos (\varphi_i - \varphi_{i+z}) \,.
\label{4}
\end{equation}
This is an excellent approximation for the original XY model for
$T \ll J_\parallel$. 
The helicity modulus parallel
to the layers is constant and given by $\Upsilon_\parallel = J_\parallel$.
In order to calculate $\Upsilon_\perp$
we treat the nonlinear term using the Villain approximation \cite{Vil75},
which is very accurate both at low $(T \ll J_\perp)$ 
and at high temperatures $(T \gg J_\perp)$ \cite{Kle89}.
The partition function is factorized into a term
representing the in-plane harmonic fluctuations and the "fluxon
contribution"
\begin{equation}
Z_{fl} = \sum_{\{m_i\}} \,e^{-2\pi^2\beta \sum_{i,j} m_iV_{ij} m_j} \, ,
\label{5}
\end{equation}
where the variables $m_i$ are integers. The interaction $V_{ij}$
is the Fourier transform of
\begin{equation}
V({\bf q}) = J_\perp^*  \, \frac{2 - \cos q_x - \cos q_y}
{2 - \cos q_x - \cos q_y + (J_\perp^* / J_\parallel) (1 - \cos q_z)}  \;  ,
\label{6}
\end{equation}
where the effective interlayer coupling is given by
\begin{equation}
J_\perp^* = 
\left( 2 \beta \,\makebox{log}\, \frac{2}{\beta J_\perp}\right) ^{-1}
\label{7}
\end{equation}
for $\beta J_\perp \ll 1$.
We notice that $V_{ij}$ is exactly equal to the interaction
energy of two elementary fluxons calculated in the usual vortex
loop representation of the 3d XY model \cite{Sav78}. The variables $m_i$
are the quantum numbers of the fluxons, and large loops can be constructed
by adding elementary fluxons.
Since the energy scale $J_\perp^*$ increases roughly linearly with
temperature the multiplication of fluxons is strongly slowed down.
The helicity modulus perpendicular to the layers, $\Upsilon_\perp$,
can be expressed in terms of fluxon variables,
\begin{equation}
\Upsilon_\perp = J_\perp^* \left\{1 - 4 \pi^2 \beta J_\perp^*
\frac{1}{N} \left\langle\left(\sum_i m_i \right)^2 \right\rangle\right\}
\,  .
\label{8}
\end{equation}
This confirms that for this approximate model the proliferation of
fluxons is directly related to the loss of interlayer coherence. 
Eq. (\ref{8}) could in
principle be used for calculating $\Upsilon_\perp$, but since the couplings
$V_{ij}$ decrease only slowly with distance (like a dipole-dipole interaction)
we have calculated $\Upsilon_\perp$ starting
from the original expression (\ref{4}), using both the RHA and Monte
Carlo simulations. The results, shown in Fig.2, demonstrate that
even for small couplings $J_\perp$ 
the helicity modulus $\Upsilon_\perp$ vanishes
only far above the BKT transition, i.e. in a temperature region where the 
Hamiltonian (\ref{4}) is
no longer a good approximation for the original XY
model. Nevertheless, the model defined by Eq. (\ref{4}) is interesting,
since it does show a layer decoupling without simultaneous loss
of intraplane coherence.
The transition temperature $T^\ast$ 
for infinitesimal interplane coupling $J_\perp$ can be 
easily calculated from the perturbation expansion described above.
In the present case the correlation function $c(r)$ decays algebraically
with an exponent $\eta(T) = T / (2 \pi J_\parallel)$, and $T^\ast$ is simply
given by the relation $\eta(T^\ast) = 1$. Fig.2 shows that the resulting
value $T^\ast = 2\pi J_\parallel$ is consistent with the Monte Carlo data.
The vanishing of  $\Upsilon_\perp$ for $T > T^\ast$ implies that also
$\langle \cos(\varphi_i) \rangle$ vanishes, although the
susceptibility remains infinite up to $2 T^\ast$.

The analysis given above strongly supports
the view  \cite{Kor90} that the interlayer coherence
is not destroyed by thermal fluctuations below the critical temperature
in the 3d XY model, even for arbitrarily small interlayer coupling.
We now address the question whether this conclusion is generally
valid, by studying
two modifications of the original XY model. The first represents
a stack of layers with periodically varying {\it intra}layer couplings.
The extensively studied superlattices of low and high $T_c$ layers
 are nice realizations of such a model
\cite{Tri94}.
The second modification concerns the {\it inter}layer coupling, which is
replaced by a different form, namely $J_\perp \cos (\varphi_i - \varphi_{i+z})$
is substituted by $(J_\perp / n^2) \cos [n (\varphi_i - \varphi_{i+z})]$,
where $n$ is an arbitrary 
positive integer $(\ge 2)$. Although such a term
cannot be excluded within Ginzburg-Landau theory,
it would in general be expected
to be much smaller than the conventional term with $n=1$.

We consider first a superlattice consisting of 
one "strong" layer with intralayer coupling $J_{\parallel}^{(1)}$, alternating
with $n$ "weak" layers with  $J_{\parallel}^{(2)} < J_{\parallel}^{(1)}$,
and a constant interlayer coupling $J_\perp$.
The Monte Carlo results  (Fig.3)
show that nothing spectacular happens for $n = 1$.
For $n = 3$, however,
the helicity modulus $\Upsilon_\parallel$ exhibits a kink slightly
above the low $T_c (\approx J_\parallel^{(2)})$, while simultaneously 
$\Upsilon_\perp$  drops practically to zero. There is apparently
a region with vanishing interlayer coherence
between this temperature and the critical temperature
$T_c \approx J_\parallel^{(1)}$.
However, a true decoupling is very
unlikely. In the extreme situation where $J_{\parallel}^{(2)} = 0$.
one can integrate out the variables of the weak layers
and deduce a model involving only the strong layers with
an effective interlayer coupling $J_\perp^{eff} \approx
(\beta J_\perp/2)^n J_\perp$.  We can then
use our previous results for one type of layers
and conclude that the helicity modulus
$\Upsilon_\perp$ remains finite (though very small) up to $T_c$.
Thus a Friedel transition does not occur for this type of superlattices.

We turn now to the second modification 
where the interlayer interaction is replaced by
$(J_\perp / n^2) \cos[n\,(\varphi_i - \varphi_{i+z})]$.
Monte Carlo results for $n = 2,3,4$ are shown in
Fig. 4. For $n > 2$ the helicity modulus $\Upsilon_\perp$
drops to zero far below the critical temperature where $\Upsilon_\parallel$
vanishes, while for $n = 2$ both transitions seem to occur at the same
temperature.
This observation is in agreement with the 
analysis of perturbation theory.
In the present situation we have to consider the correlation function
\begin{equation}
c_n(r) =\langle \cos[n\,(\varphi_i - \varphi_{i+r})] \rangle \sim
r^{-\eta_n(T)}
\label{9}
\end{equation}
For the exponent $\eta_n$ we use the relation \cite{Jos77} 
$\eta_n = n^2\,\eta$
 together with the numerical values for $\eta(T)$
\cite{All88}. The criterion $\eta_n(T^\ast) = 1$ yields decoupling
temperatures $T_n^*$ in good agreement with those found in the
Monte Carlo simulations for $n = 3,4$.  For $n=2$
the reported value $\eta(T_{KT}) = 1/4$ indicates
that both transitions --
the Friedel and the BKT transition -- occur simultaneously.
 
In summary, our numerical simulations and perturbative arguments
confirm that for the layered XY model an intermediate effectively
$2d$ phase does not exist, even for arbitrarily small interlayer
couplings, $J_\perp \ll J_\parallel$. This absence of a low-temperature
decoupling transition is nicely illustrated in a simplified version
of the model where the role of fluxons is particularly transparent.
The interplane coherence also persists for superlattices of high-
and low-$T_c$ layers, although an apparent decoupling is observed
for thick enough low-$T_c$ layers. In contrast, a low-temperature
decoupling transition is found for an interlayer coupling with
n-fold symmetry, at least for $n>2$.

We gratefully acknowledge useful discussions with P. W. Anderson,
L. N. Bulaevski, D. Feinberg, B. Horovitz, S. E. Korshunov and
S. Shenoy.
This work has been supported by the Swiss National 
Foundation through grant Nos. 4030-32799 and 20-40672.94.

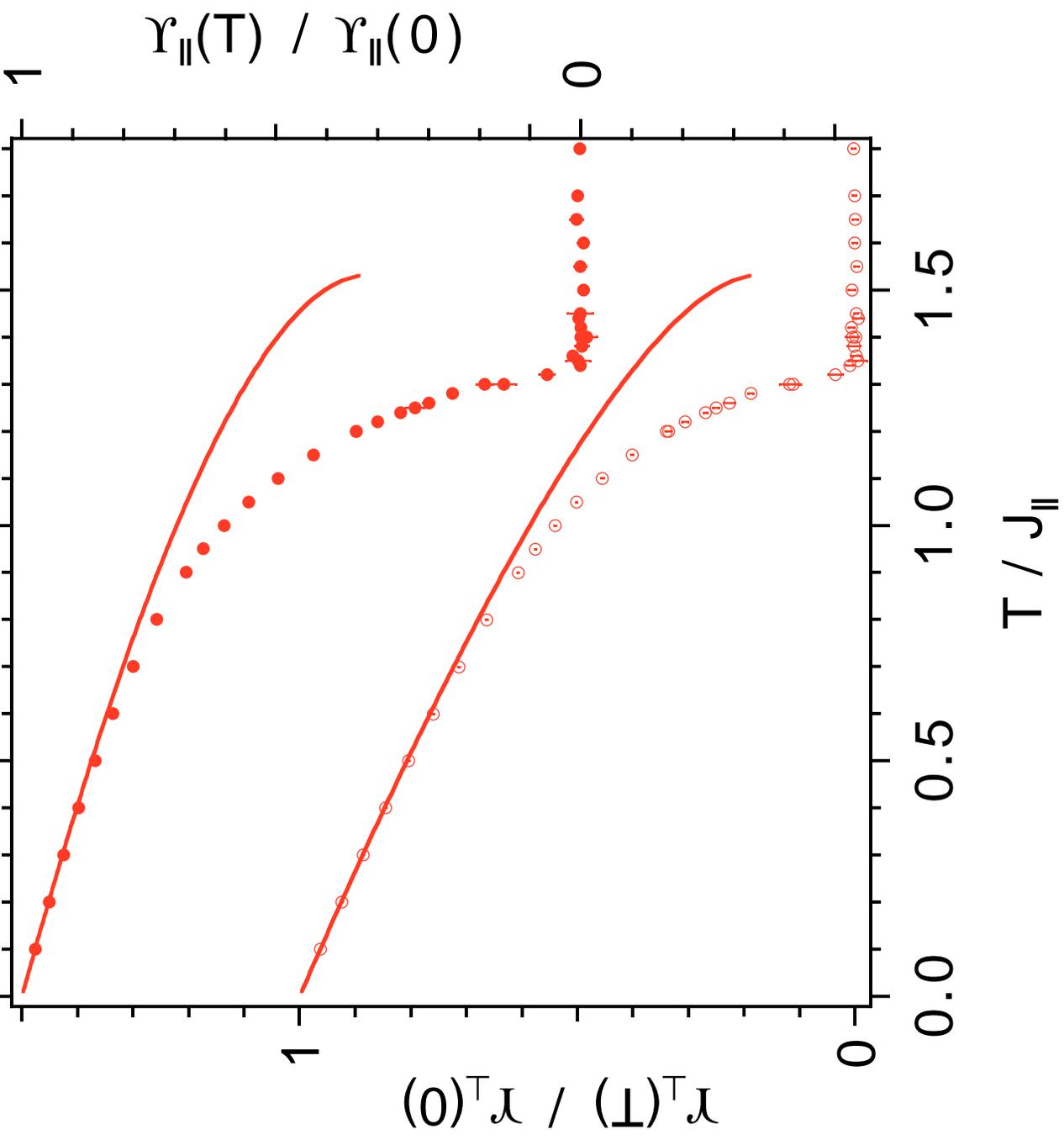
\begin{figure}
\caption{Helicity moduli $\Upsilon_\parallel$ (full circles) and
$\Upsilon_\perp$ (open circles)
of the anisotropic XY model with $J_\perp / J_\parallel = 0.1$
as functions of temperature
as obtained from Monte Carlo simulations of a 
$36\times 36\times 36$ lattice.
The solid curves show the RHA results.}
\label{fig1}
\end{figure}

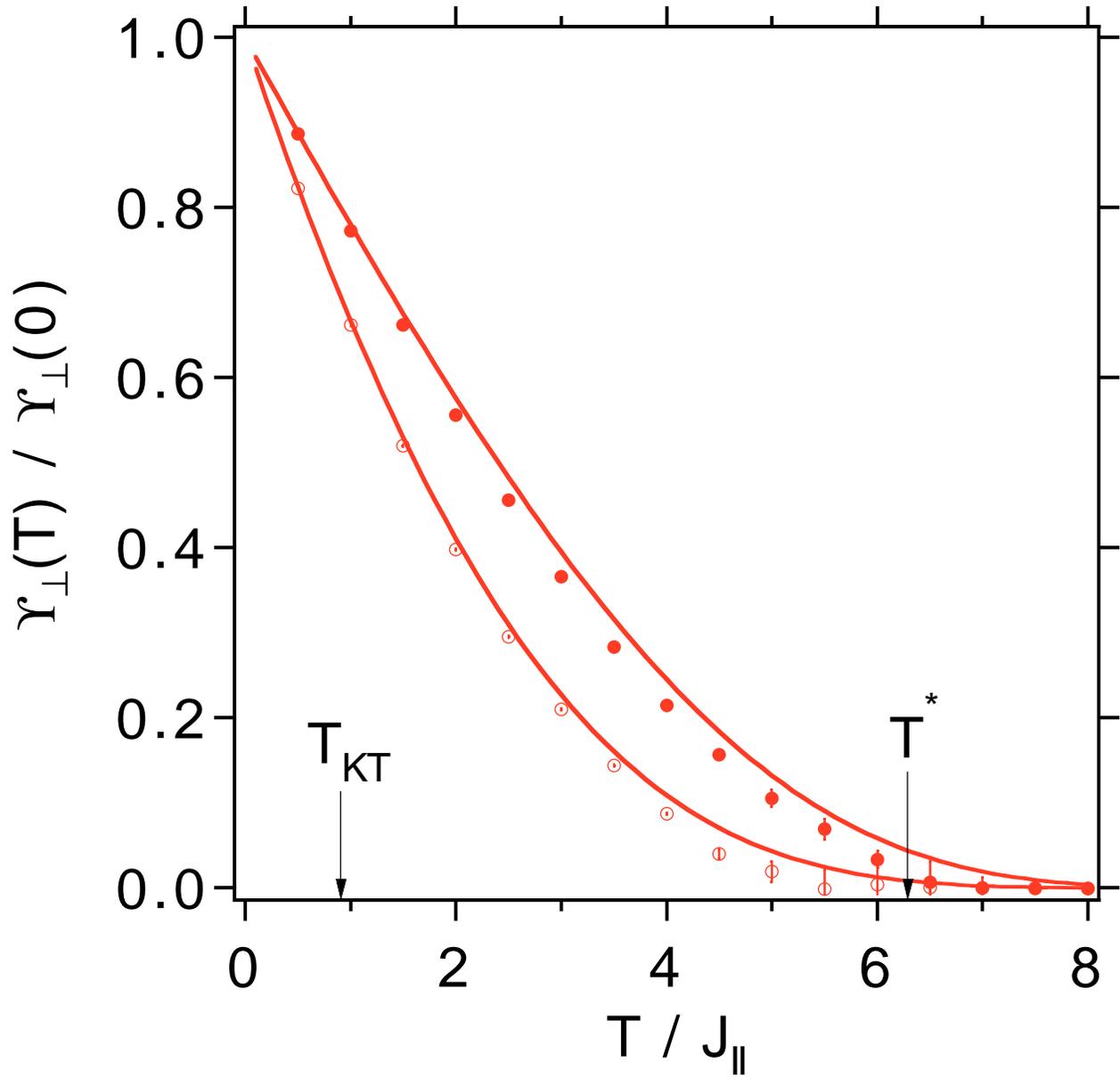
\begin{figure}
\caption{Helicity modulus $\Upsilon_\perp$ of the approximate version
(Eq. (4)) of the XY model with $J_\perp / J_\parallel = 0.1$ (open circles)
and $J_\perp / J_\parallel = 0.5$ (full circles). The symbols are Monte Carlo
data and the full lines are RHA results.
Arrows indicate the Kosterlitz--Thouless temperature $T_{KT}$ and
the layer decoupling
temperature $T^\ast = 2 \pi J_\parallel$, respectively. }
\label{fig2}
\end{figure}

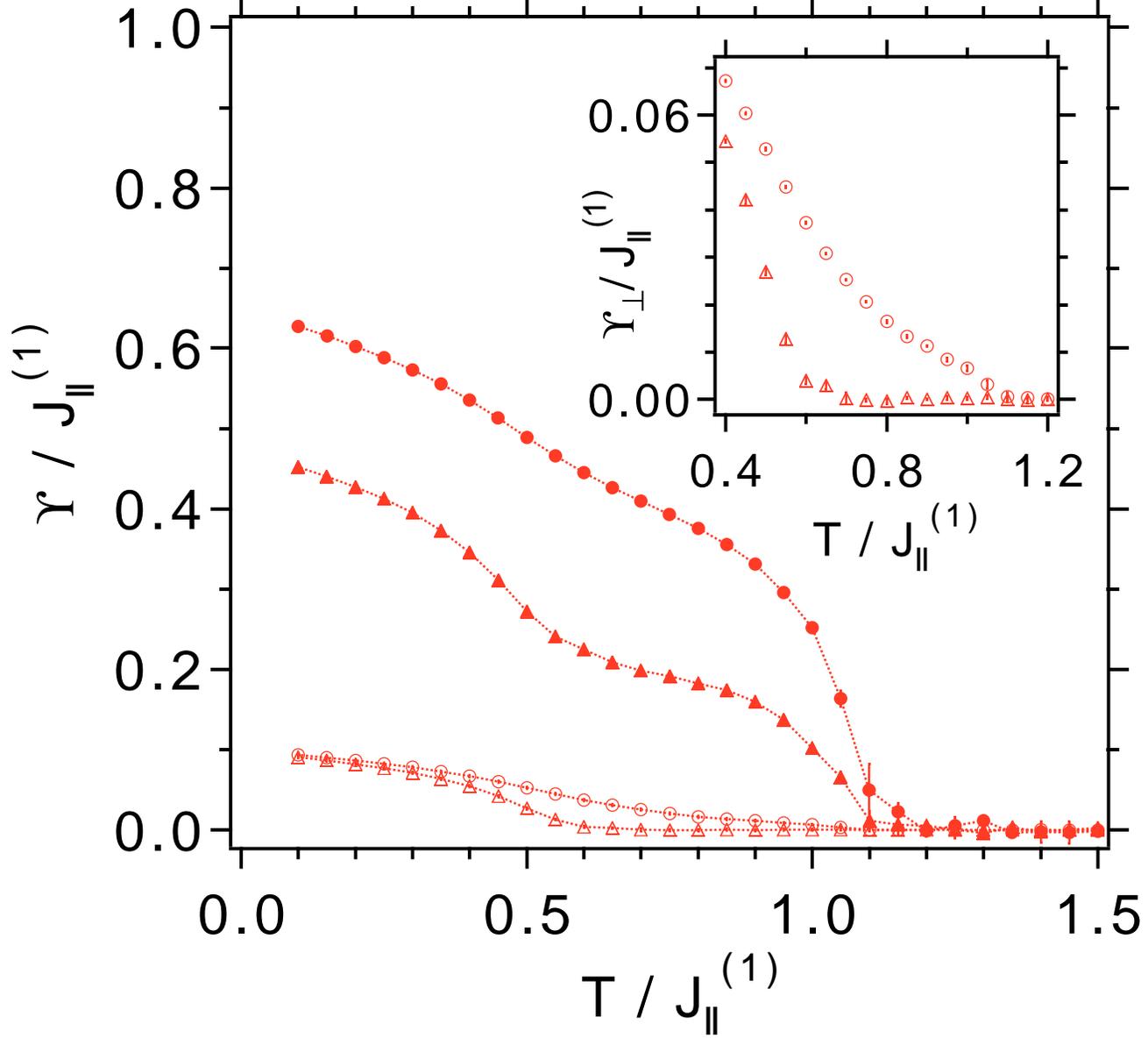
\begin{figure}
\caption{Helicity moduli $\Upsilon_\parallel$ (full symbols) and
$\Upsilon_\perp$ (open symbols)
for the superlattice model (as defined in the text) 
with   $J_\parallel^{(2)} / J_\parallel^{(1)} = 0.3$
and $J_\perp / J_\parallel^{(1)} = 0.1$.
Monte Carlo data for a $36\times 36\times 36$ lattice with one (circles) and
three  (triangles) weak layers are presented.
The inset shows $\Upsilon_\perp$ in the crossover region.}
\label{fig3}
\end{figure}

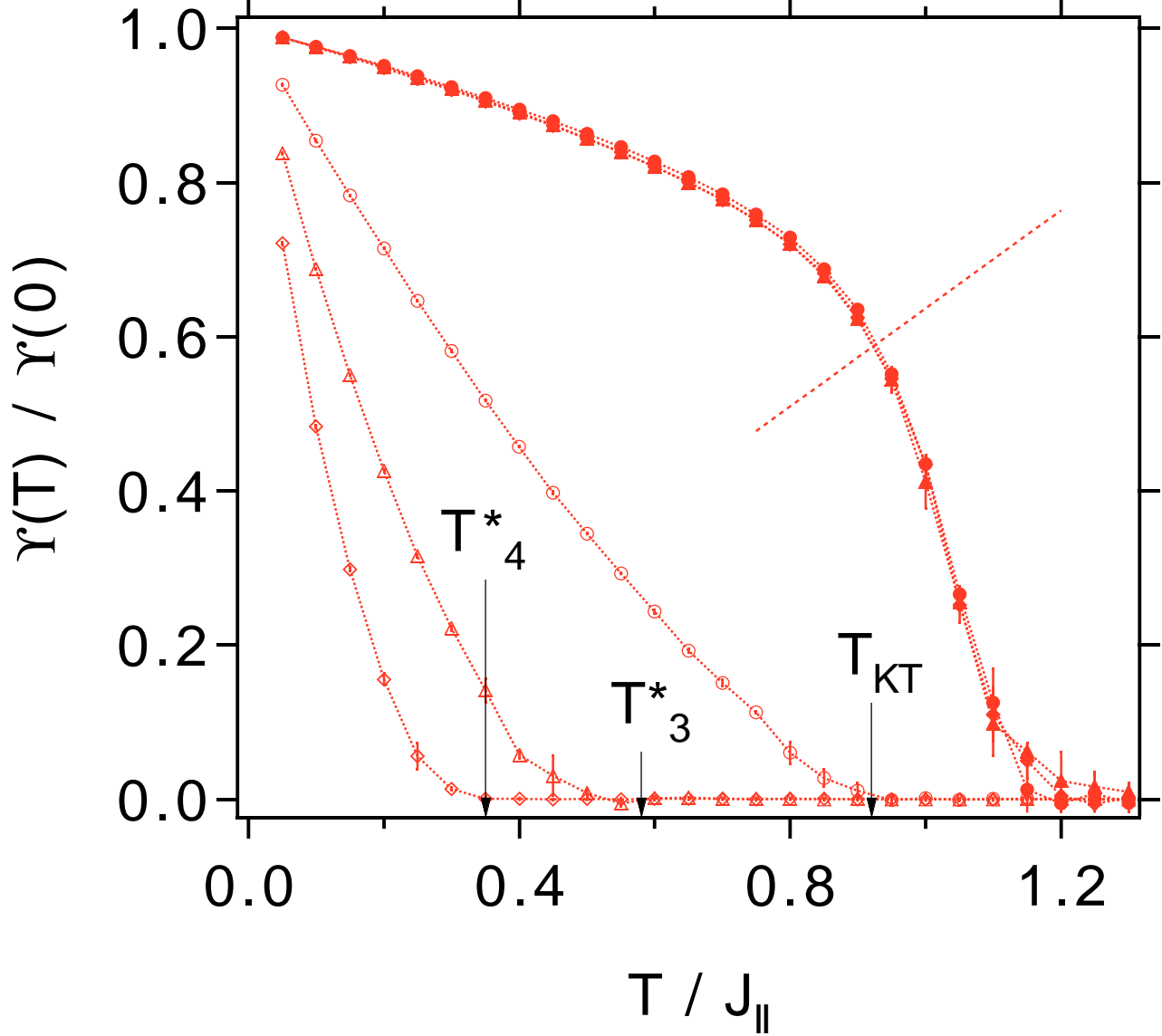
\begin{figure}
\caption {Helicity moduli $\Upsilon_\parallel$ (full symbols) and
$\Upsilon_\perp$ (open symbols) of the  XY model 
with modified interplane interaction 
$(J_\perp / n^2) \cos[n\,(\varphi_i - \varphi_{i+z})]$
with $J_\perp / J_\parallel = 0.1$.
Monte Carlo data for $n = 2$ (circles),
$n = 3$ (triangles) and $n = 4$ (diamonds) are shown.
The intersection of the 
dashed line with slope $2 / \pi$ and the $\Upsilon_\parallel$ curves 
locates the BKT transition at $T_{KT}$.
Arrows indicate the calculated layer decoupling temperatures 
$T^\ast_n$ for $n = 3,4$.}
\label{fig4}
\end{figure}


\begin{references}
\bibitem{Atthi} At this point we do not know a physical realization
of this model.
\end{references}
\end{document}